\documentclass[graybox, envcountchap]{svmult}

\usepackage{mathptmx}        
\usepackage{helvet}          
\usepackage{courier}         

\usepackage{makeidx}         
\usepackage{graphicx}        
\usepackage{multicol}        
\usepackage[bottom]{footmisc}

\makeindex             

\begin{document}


\title*{Gravitational waves from isolated neutron stars}
\author{Brynmor Haskell and Kai Schwenzer}
\institute{Brynmor Haskell \at Nicolaus Copernicus Astronomical Center, Polish Academy of Sciences, Bartycka 18, 00-716 Warsaw, Poland, \email{bhaskell@camk.edu.pl}
\and Kai Schwenzer \at Istanbul University, Science Faculty, Department of Astronomy and Space Sciences, Beyaz\i t, 34119, Istanbul, Turkey, \email{kai.schwenzer@istanbul.edu.tr}}
%
%
\maketitle

\abstract{Neutron star interiors are a fantastic laboratory for high density physics in extreme environments. Probing this system with standard electromagnetic observations is, however, a challenging endeavour, as the radiation tends to be scattered by the outer layers and the interstellar medium. Gravitational waves, on the other hand, while challenging to detect, interact weakly with matter and are likely to carry a clean imprint of the high density interior of the star. In particular long lived, i.e. `continuous' signals from isolated neutron stars can carry a signature of deformations, possibly in crystalline exotic layers of the core, or allow to study modes of oscillation, thus performing gravitational wave asteroseismology of neutron star interiors. In this article we will review current theoretical models for continuous gravitational wave emission, and observational constraints, both electromagnetic and gravitational. Finally we will discuss future observational possibilities.}

\section{Introduction}

Once detected, continuous gravitational wave (GW) radiation could provide a novel window to the cosmos that would perfectly extend our present electromagnetic spectrum to sub-kHz frequencies. In contrast to observed transient gravitational wave events \cite{GBM:2017lvd}, these would more closely resemble ordinary astronomical sources in that they would allow us to perform precision long-term observations that could probe the presently inaccessible interior of dense compact objects \cite{Haenselbook}. In this chapter we give a short overview of this promising subject, ranging from their theoretical description, the properties of the compact sources that emit them and possible emission mechanisms, as deformations and oscillation modes, to the indirect impact of such an emission on electromagnetic multi-messenger observations, as well as current searches and the direct constraints from their present non-detection.

\section{Continuous gravitational wave emission from rotating sources}

The creation of gravitational waves, of a size that could potentially be detected, in general requires huge masses moving with large velocities in an asymmetric way. These requirements are presently only fulfilled for compact sources, like black holes and neutron stars. Whereas featureless black holes cannot produce a sustained emission over longer time intervals, in neutron stars there are known processes that could power a continuous gravitational wave emission.

Since compact sources are very degenerate systems close to the groundstate of matter, dynamical processes that could cause a gravitational wave emission are mostly absent. There are then basically two possibilities for continuous gravitational wave emission, namely that the star is statically, asymmetrically deformed and the rotation causes a time dependent quadrupole moment, or there are oscillation modes of the star \cite{Kokkotas:1999bd} that are unstable to gravitational wave emission. However, dramatic events like the core bounce at birth, special events during the evolution of the source, e.g. phase transitions, crust quakes and glitches, accretion, or tidal deformations before merger, can excite transient oscillations that could be detectable. 

The main modes that can become unstable in compact sources are the fundamental (f-) modes and the Rosby (r-) modes. F-modes are polar modes, whereas r-modes are axial modes, that are only present in rotating stars. They are obtained from a solution of the perturbed hydrostatic equation in a slow rotation expansion. R-modes to leading order in a Newtonian approximation do not involve compressions of the star and are given by a simple analytic expression for the velocity perturbation, independent of the detailed structure of the source \cite{Lindblom:1998wf}
\begin{equation}
\delta\vec{v}=\alpha R\Omega\left(\frac{r}{R}\right)^{m}\vec{Y}_{mm}^{B}\!\left(\theta,\varphi\right)\mathrm{e}^{i\omega t} \label{eq:r-mode}
\end{equation}
where $R$ is the radius of the source, $\vec Y^{B}$ are magnetic vector spherical harmonics and $\alpha$ is a dimensionless amplitude parameter. Moreover, they have in this approximation a fixed canonical relation between oscillation and rotation angular velocity $\omega =4/3 \Omega$, although note that there will be corrections to this relation due to rotation and general relativity.

\subsection{Multipole radiation} 

Since there is no way to observationally resolve compact objects they present perfect point sources. Correspondingly the gravitational wave emission can be treated within a multipole expansion in terms of the leading coefficients. Gravitational waves are described by the deviation of the metric from its flat Minkowski form.
The general expression for the transverse traceless deviation reads \cite{Thorne:1980ru}
\begin{equation}
h^{TT}_{ij}=\frac{G}{c^4 r} \sum_{i=2}^\infty \sum_{j=-l}^l \left( \frac{d^l}{(dt)^l} I_{lm} (t-r) T^{E2}_{lm,ij} + \frac{d^l}{(dt)^l} S_{lm} (t-r) T^{B2}_{lm,ij}\right)
\end{equation}
where $T^{E2}$ and $T^{B2}$ are electric and magnetic tensor spherical harmonics and $I$ and $S$ are generalized mass and current multipole moments.

The continuous gravitational wave radiation from compact sources typically has a quasi-periodic form with time dependence $\exp ((i\omega-1/\tau)t)$, where $\omega$ is the angular velocity and $\tau$ the driving/damping time. Generally, $2\pi/\omega \ll \tau$, so that a harmonic form for the gravitational wave is sufficient and the evolution of the signal on a secular time scale $\tau$ only affects the amplitude of the gravitational wave signal and can be included subsequently, so that the multipole coefficients reduce to conventional time-independent multipole moments.

The lowest of these modes generally dominates the gravitational wave emission. The relevant mass quadrupole moments for gravitational wave emission due to deformations (or `mountains') of the star and f-modes as well as the current quadrupole moments for r-mode oscillations are for sufficiently slowly moving sources \cite{Thorne:1980ru}
\begin{eqnarray}
I_{2m}&=&\frac{16\sqrt{3}\pi}{15}\int \tau_{00} Y_{2m,j}^{*}r^2 d^{3}x \\
S_{22}&=&\frac{32\sqrt{2}\pi}{15}\int (-\tau_{0j}) Y_{22,j}^{B*}r^2 d^{3}x 
\end{eqnarray}
respectively, where $\tau$ is the stress-energy tensor, $Y$ are scalar and $Y^{B}$ magnetic vector spherical harmonics.

\subsection{Gravitational wave strain}

As a measure for the detectability of gravitational wave signals usually the intrinsic gravitational wave strain amplitude $h_0$ is used. It represents the amplitude that would be measured in the idealized case that a detector is positioned at one of the earths poles and observes a source vertically above it, whose rotation axis is parallel to that of the earth. 

For gravitational waves due to deformations of the star the intrinsic strain amplitude reads \cite{Owen:2010ng}:
\begin{equation}
    h_{0}=\sum_{m=-2,2} 4\pi^2 \frac{G}{c^4}\sqrt{\frac{5}{8\pi}}\frac{f^2}{D}|I_{2m}|
\end{equation}
although note that in observational gravitational wave papers a slightly different definition of the mass quadrupole $Q_{22}=\int \Re{(\delta\rho_{22})} r^2 dr$ is commonly used, with $\delta \rho_{22}$ the $l=m=2$ component of the density distribution, such that
\begin{equation}
    h_{0}=4\pi^2 \frac{G}{c^4}\sqrt{\frac{8\pi}{15}}\frac{f^2}{D} Q_{22}
\end{equation}
where $f$ is the frequency of the wave (for a pulsar rotating at a frequency $\nu$, $f=2\nu$), $D$ the distance to the source and $I_{zz}$ the relevant component of the moment of inertia tensor.

Often, for `mountains', this result is presented in terms of an ellipticity $\epsilon$, defined as:
\begin{equation}
 \epsilon=\frac{I_{xx}-I_{yy}}{I_{zz}}=\frac{Q_{22}}{I}\sqrt{\frac{8\pi}{15}}\, ,  
\label{epsilondef} \end{equation}
where $I_{ii}$ are the principal moments of inertia and $I_{xx}\approx I_{yy} \approx I_{zz}\approx I$.
One can then simply write:
\begin{equation}
    h_0=4\pi^2\frac{G}{c^4}\frac{f^2}{D}\epsilon
\end{equation}

For gravitational waves due to classical r-modes eq.~(\ref{eq:r-mode}) the current quadrupole moment reduces to
\begin{equation}
S_{22}=\frac{32\sqrt{2}\pi}{15c}\tilde{J}M R^{3}\Omega\alpha
\end{equation}
where the dimensionless constant $\tilde J$ characterizes the source \cite{Owen:1998xg} via an integral over the energy density $\rho$
\begin{equation}
\tilde{J}\equiv\frac{1}{MR^{2m}}\int_{0}^{R}dr\, r^6 \rho
\end{equation}
The relation between the oscillation and rotation angular velocity can be written
\begin{equation}
\omega = \left( \kappa(\Omega)-2 \right) \Omega \equiv - \frac{4}{3} \chi (\Omega) \Omega= -\frac{2\chi}{3\pi}  f
\end{equation}
where $\chi \approx 1$ parametrizes the deviation from the canonical relation due to rotation and general-relativistic corrections. The intrinsic strain amplitude takes then form 
\cite{Kokkotas:2015gea}
\begin{equation}
h_{0}=\sqrt{\frac{2^{9}}{3^{6}5\pi^5}} \frac{G}{c^5} \tilde{J}MR^{3} \frac{\chi^{2} f^{3}\alpha}{D}    
\end{equation}

The likelihood of detecting continuous gravitational waves in a particular search obviously increases with the observation time $\Delta t$. This is usually taken into account by giving the equivalent intrinsic strain amplitude that would be observed within a given search at 95\% confidence level. For a coherent search it is given by
\begin{equation}
h_{0}^{95\%}=\Theta\sqrt{\frac{S_{h}\!\left(f\right)}{\Delta t}}\:.\label{eq:observational-intrinsic-strain}
\end{equation}
in terms of the power spectral density $S_h$ of the noise and a factor $\Theta$ that depends on the particular search ($\Theta\approx11.4$ for known pulsars  and $\Theta\approx 35$ for sources without timing data \cite{Wette:2008hg}).

\section{Composition and material properties of compact stars}
\subsection{Phases of dense matter}
\label{dense}
Continuous gravitational waves are very promising as probes for the otherwise inaccessible interior of compact sources \cite{Haenselbook}, whose composition is still largely unknown. Even though the theory of matter, quantum chromodynamics (QCD), is known and intensely studied \cite{Brambilla:2014jmp}, the particular phase(s) realized at compact star densities is not. There are dozens of proposed potential phases and countless further uncharted possibilities \cite{Alford:2007xm}. Yet, they form two qualitatively distinct classes, namely forms of hadronic matter composed of the familiar hadronic particles (neutrons, protons, ...), that also make up ordinary matter, and deconfined quark matter, whose long-distance carriers of conserved quantum numbers are the underlying colored quark constituents. I.e. in the latter case quarks are not confined and dynamical mass generation due to non-perturbative QCD effects, which is responsible for $95\%$ of the mass of atomic matter, could be dramatically reduced. 

The simplest and prototypical case for the composition of the interior, responsible for the term "neutron star", is npe$\mu$-matter, i.e. a liquid of nearly $90\%$ neutrons and around $10\%$ protons as well as an equal number of leptons to electrically neutralize the star \cite{Haenselbook}. Similarly on the quark side the simplest option is strange quark matter consisting of u, d and s quarks neutralized by a small fraction of electrons. The first form of matter is expected at sufficiently low densities, whereas forms of quark matter are theoretically predicted to be realized at asymptotically large densities that are significantly above the density range accessible in the core of compact stars. In this regime the novel phenomenon of color-superconductivity \cite{Alford:2007xm} should strongly determine the properties of the corresponding matter. In both limits there are controlled approximation schemes, namely chiral perturbation theory and the weak coupling expansion owing to the asymptotic freedom of QCD \cite{Brambilla:2014jmp} and these can to some extent constrain bulk properties in the intermediate regime. Nevertheless, the potential options for the composition and symmetry structure at realistic densities are staggering due to the large number of hadrons on the hadronic side (n, p, hyperons, resonances, pions, ...) and the potential symmetry breaking patterns on the quark side. Both the presence of hadrons and the symmetry of the ground state are determined by the strong interactions in a dense medium which are still only poorly understood due to the inherent complexity to solve QCD at finite density \cite{Brambilla:2014jmp}. Gravitational wave astrophysics could therefore offer a unique way to probe the dense interior and learn about fundamental properties of matter and the realized phase(s) at neutron star densities.

\subsection{Rigidity and shear modulus}

The exterior layers of the neutron star, at densities below $\rho\approx 10^{14}$ g/cm$^3$, constitute the crust of the star. Once the star has cooled to temperatures below a few times $10^9$ K, after approximately a minute from birth, the crust solidifies and forms a crystal structure. The composition of the nuclei in the crystal depends on density, with increasingly neutron rich nuclei appearing at higher densities (see \cite{Haenselbook} for an in depth discussion).

For the present discussion we are mainly interested in the crust's rigidity, as this determines to what extent it can support a deformation, and eventually a GW emitting `mountain'. This information is encoded in the shear modulus, which can be estimated with Monte-Carlo simulations of deformations, assuming the crust forms a body-centred cubic (BCC) lattice, leading to \cite{NilsBook}:
\begin{equation}
    \mu\approx 10^{20} \left(\frac{\rho}{10^{14}\,\mbox{g cm$^{-3}$}}\right)\,\mbox{g cm$^{-1}$ s$^{-2}$}\, .
\end{equation}

Compared to the pressure the shear modulus is weak, so the crust behaves more like a jelly than a traditional solid. To understand it's relevance for gravitational wave emission, however, another quantity is important, and this is the breaking strain $\sigma_b$. The breaking strain quantifies the level of strain, defined for example by the Von Mises criteria as $\sigma=\sqrt{t_{ab}t^{ab}}$ with $t_{ab}$ the strain tensor, above which the crust will yield. For terrestrial materials the breaking strain lies roughly in the range $10^{-5}< \sigma_{br}<10^{-2}$, but molecular dynamics simulations suggest $\sigma_{br}\approx 0.1$ for a NS crust \cite{HorowitzKadau}, which allows for a substantial `mountain' to build up and emit gravitational waves, as we shall see.

Before moving on, let us remark that many equations of state predict `pasta' phases, in which nuclei form rods and plates and other structures, at higher densities close to the base of the crust. In this case one would have a structure similar to a liquid crystal, which a much reduced shear modulus, and possibly breaking strain \cite{NilsBook}. Similarly there may be crystalline phases of quark matter in the deep core, which stem from a spatially periodic fluctuation of the color-superconducting condensate \cite{Alford:2007xm}. Since the critical energy of color-superconductivity is much larger than typical nuclear binding energies in the hadronic crust, this "solid" is much more rigid, can resist shear forces and could support large `mountains' that would be detectable by ground based GW detectors.

\subsection{Viscous damping}

Dynamic material properties, like dissipation, can differ dramatically depending on the form of matter that is present inside a neutron star. The reason for this is that neutron stars are highly degenerate systems where low energy processes are parametrically suppressed and depend sensitively on the degrees of freedom and the interactions between them. The dissipation of large scale motion is described by the viscosity coefficients in a hydrodynamic description. In case the different microscopic degrees of freedom form approximately a single non-relativistic fluid, the dissipated energy density is to leading order described by the shear viscosity $\eta$ and the bulk viscosity $\zeta$
\begin{equation}
\frac{d\epsilon}{dt}\approx- \eta\left(\partial_{i}v_{j}\right)^{2}-\left(\zeta+\frac{1}{3}\eta\right)\left(\partial_{i}v_{i}\right)^{2}
\end{equation}
while more coefficients arise in relativistic, asymmetric or multi-fluid cases \cite{Romatschke:2017ejr}.

The viscosity coefficients describe the dissipative momentum transfer in the fluid and are obtained from the underlying kinetic theory \cite{Romatschke:2017ejr}. The shear viscosity stems from friction due to shear motion in the fluid itself or along a boundary \cite{Lindblom:2000gu}. Like most transport properties it is dominated by the particle with the largest mean free path (i.e. being the lightest and/or having the weakest interactions) and it is the larger, the longer the latter is. In npe$\mu$-matter it is at neutron star temperatures generally dominated by leptons with long-ranged electromagnetic interactions \cite{Shternin:2008es}. In a quark star, quark scattering due to the strong interaction leads to a similar but smaller result. However if fermionic degrees of freedom are absent due to Cooper pairing, like in the CFL phase \cite{Alford:2007xm}, the shear viscosity can be drastically different.

Bulk viscosity describes the dissipation in compression and rarefaction processes that push the system out of statistical equilibrium. It is a resonant processes and becomes large when the time scale of the underlying interactions matches the time scale of dynamic processes, like oscillations of neutron stars. In neutron stars bulk viscosity typically stems from slow weak interaction processes, causing chemical equilibration. For small deviation $\delta\mu$ of the chemical potential from chemical equilibrium $\delta \mu \ll T$ it has the general form \cite{Alford:2010gw}
\begin{equation}
\zeta=\frac{C^{2}\tilde{\Gamma}T^{^{\delta}}}{\omega^{2}+B^{2}\tilde{\Gamma}^{2}T^{^{2\delta}}} \label{eq:bulk-viscosity}
\end{equation}
Here $\Gamma=\tilde\Gamma T^\delta \delta\mu$ is the weak equilibration rate and $B$ and $C$ are susceptibilities determining the flavor dependence of the equation of state of dense matter. As seen in eq.~(\ref{eq:bulk-viscosity}), the resonant maximum is tuned by the equilibration rate and therefore for a given angular frequency $\omega$ reached at rather different temperatures for different forms of matter. In npe$\mu$-matter chemical equilibration proceeds via semi-leptonic Urca-processes. Yet, except for extremely heavy stars direct processes are generally not kinematically allowed and only slow modified Urca reactions are possible, where additional strong interactions guarantee momentum conservation. At neutron star temperatures the dissipation is therefore far from the resonant maximum. In strange hadronic or quark matter in contrast also non-leptonic processes are possible which have a much larger rate, leading to a resonant maximum of the dissipation at temperatures present in neutron stars. The absence of low energy fermionic modes again strongly alters the dissipation \cite{Alford:2007xm}. In phases separated by a sharp phase boundary, there is likewise an analogous boundary version of bulk viscosity stemming from phase conversion at the boundary.

\subsection{Heat transport and cooling}

The dissipation of large scale fluid motion can strongly heat the star. This heat which is generally non-uniformly created throughout the star has to be transported to ensure thermal equilibrium and eventually has to be radiatively lost again \cite{Haenselbook}. Therefore, also thermal properties of dense matter are important to assess the evolution and gravitational wave emission of continuous sources. The thermal conductivity, dominated by weakly interacting, relativistic fermionic particles is typically very large in neutron stars unless degenerate fermions are absent. In npe$\mu$-matter it is the electrons that dominate heat transport and in strange quark matter the quarks. Correspondingly the core of such stars is typically in thermal equilibrium.

Unless the temperature becomes very low, the cooling of the star is dominated by neutrino emission from the bulk of the star. If fermions are present these are dominated by the same Urca processes that also cause bulk dissipation. In neutron matter the modified Urca processes lead to slow cooling of the star \cite{CoolReview}.
In addition the neutrino emissivity in superfluid and superconducting phases is enhanced close to the critical temperature. Finally, direct Urca processes at very high density present the fastest known cooling mechanism \cite{CoolReview}. The latter are always present in quark matter so stars containing quarks would typically be significantly cooler than hadronic stars.  

\section{Gravitational waves due to `mountains'}

The rigidity of the crust, discussed in the previous sections, is a key ingredient for one of the first proposed mechanisms to create a time varying quadrupole on a neutron star. The idea is that the crust can sustain a `mountain' on the neutron star in much the same way as the terrestrial crust sustains mountains on Earth. Of course, due to the much larger gravitational pull, neutron star mountains are much smaller, with typical estimates of the heights of the order of $\delta r\approx 1$ mm \cite{NilsBook}. In a gravitational wave context, however, we have seen that the size of the mountain is usually discussed in terms of its ellipticity, defined in (\ref{epsilondef}) assuming that the $l=m=2$ multipole dominates the emission.
In this case the GW emission will be at twice the rotation frequency of the star, and the GW amplitude is 
\begin{equation}
h_0\approx 4\times 10^{-24} \left(\frac{\mbox{ms}}{P}\right)^2 \left(\frac{\mbox{kpc}}{r}\right)\left(\frac{I}{10^{45} \mbox{g cm$^2$}}\right)\left(\frac{\epsilon}{10^{-6}}\right)\,\label{strainc}
\end{equation}
with $r$ the distance to the source and $P$ the period of the star.

Furthermore, the gravitational radiation carries away angular momentum, and spins down the star at a rate
\begin{equation}
    \dot{\nu}\approx-2.67\times 10^{-8} \left(\frac{\mbox{ms}}{P}\right)^5\left(\frac{I}{10^{45} \mbox{g cm$^2$}}\right)\left(\frac{\epsilon}{10^{-6}}\right)^2\, ,
\label{spindown} \end{equation}
which leads to a braking index $n=\ddot{\nu}\nu/\dot{\nu}^2=5$.

Note, however, that in the more general case where the star is a triaxial ellipsoid that is not rotating around one of its principal axis of inertia, one expects emission also in the $l=2, m=1$ harmonic, which leads to emission at the same frequency as the stellar rotation. The main theoretical models for such a scenario are precessing neutron stars and, as we shall see in the following, deformed magnetised neutron stars.

Mountains can also be supported by the strong magnetic field of the NSs, which deforms the spherical shape of the star and, in fact, provides a lower limit to the size of the quadrupole. In accreting neutron stars the flow of matter can bury the field and locally enhance it's strength and confine a larger mountain, as well as source reactions deeper in the crust that can lead to deformations, as we will see in the following.

\subsection{Crustal mountains}

As already mentioned the neutron star crust can support shear strains and thus a `mountain', where in this section we will use this term generally to indicate a quadrupolar deformation of the star (naturally the overall angular structure of the mountain may be more complex, but the quadrupolar part will give the strongest contribution to the GW emission). In general mature NSs are expected to be almost perfectly spherical, as during the life of the star processes such as plastic flow and crustquakes are expected to release the strain in the crust and erase any mountain.

Things are, however, different for neutron stars in Low Mass X-ray Binaries (LMXBs) that are accreting from a less evolved companion, and being spun up in the process.
In this case light elements are being accreted onto the NS from the companion and slowly compressed to higher densities, where they undergo a number of reactions such as electron captures and pycnonuclear reactions \cite{Haenselbook} which release energy locally and heat up the crust, in a process known as `deep crustal heating'. When accretion outbursts end, the neutron star crust can be observed to cool, by observing the spectrum of LMXBs entering quiescence (see \cite{CoolReview} for a review).

If the heating is not entirely symmetric, due to asymmetries in the accretion process, and there are thermal or compositional gradients, this can source deformations of the crust and lead to a quadrupole, which can be estimated, for a single reaction layer, as \cite{UCB2000}:
\begin{equation}
Q_{22}=3\times 10^{35} R_{12}^4 \left(\frac{\delta T_q}{10^5\mbox{K}}\right) \left( \frac{E_{th}}{30\mbox{MeV}}\right)^3\,\mbox{g cm$^2$}\, ,
\end{equation}
where $\delta T_q$ is the $l=m=2$ part of the temperature increase due to deep crustal heating, $R_{12}$ the radius of the star normalised to 12 km, and $E_{th}$ the threshold energy for the reaction layer we are considering. For the whole crust one can estimate deformations of up to $\epsilon\approx 10^{-6}$ \cite{UCB2000}, which is within the theoretical limits of what the crust can sustain before breaking, which is of the order of $\epsilon_{max}\approx 10^{-5} (\sigma_m/0.1)$, with $\sigma_m$ the breaking stain of the NS crust, which molecular dynamics simulations find to be of order $\sigma_m\approx 0.1$, as previously discussed \cite{HorowitzKadau}.

It has, in fact been suggested that gravitational wave torques, possibly due to mountains, are setting the spin-limit of neutron stars in LMXBs \cite{UCB2000}. All the NSs in these systems are spinning well below their Keplerian breakup limit, independently of the equation of state \cite{HaskellSpin}. As the accretion torques in the system spin the stars up, additional torques must necessarily be at work, either due to the interaction between the accretion disc and the magnetic field of the star, or due to gravitational waves, from mountains or r-modes, as we will discuss in detail in the following sections.

Some systems show, in fact, tantalising evidence of a mountain, as we will discuss in the following \cite{HaskellPRL17}. Recent models have estimated the height of a mountain that can be created in the outer layers of an accreting NS, due to asymmetric accretion and consequent asymmetric heating \cite{Singh20}. In general the sizes are too small to be detected by ground based interferometers such as LIGO and Virgo, but if additional shallow heating reactions are present, as suggested also by observations of the cooling of X-ray transients \cite{CoolReview}, they may be sizable enough to explain the observed phenomenology.

\subsection{Magnetic deformations}

It is well known that magnetic stars are not spherical, as the Lorentz force leads to deformations of the stellar structure. If the magnetic axis of the star is not aligned with the rotation axis, this leads to a time varying quadrupole and gravitational wave emission. In general, given an inclination angle $\alpha$ between the magnetic and rotational axis of the star, we expect emission both in the $l=m=2$ harmonic at twice the rotation frequency, and in the $l=2, m=1$ harmonic, at the rotation frequency $\Omega$. The amplitude of the GWs scales approximately as \cite{Bonazzola96}:
\begin{equation}
h_0\approx h_{21} \sin\alpha\cos\alpha + h_{22}\sin^2\alpha\, ,  
\end{equation}
so we see that for large inclination angle emission at twice the rotation rate in the $l=m=2$ harmonic dominates, while for small angles emission at the rotation frequency is dominant and, naturally, there is no emission if $\alpha=0$ and the magnetic and rotational axes coincide, so that the deformation is axisymmetric. For the sake of definiteness we will focus on the $Q_{22}$ harmonic in the following, and discuss the related ellipticites. However the order of magnitude of $Q_{21}$ is expected to be similar and the discussion would be qualitatively equivalent. In fact, we shall see that in practice many models simply calculate the axisymmetric deformation $Q_{20}$, and assume that if the magnetic axis is inclined with respect to the rotation axis, one has $Q_{21}\approx Q_{22}$.

A large amount of work has been devoted to calculating the quadrupolar distortions of magnetised neutron stars (see \cite{LaskyReview} for a review), and the general result is that poloidal magnetic fields make the star oblate, with
\begin{equation}
\epsilon\approx 10^{-12} \left(\frac{\bar{B}}{10^{12}\mbox{G}}\right)^2\, ,
\end{equation}
with $\bar{B}$ the volume averaged magnetic field strength, while a toroidal field leads to a prolate star with
\begin{equation}
\epsilon\approx - 10^{-11} \left(\frac{\bar{B}}{10^{12}\mbox{G}}\right)^2\, .
\end{equation}
The field in a neutron star is expected to be a combination of poloidal and toroidal components, a so-called 'twisted torus' in which the poloidal component that threads the star and extends to the exterior is stabilised by a weaker internal toroidal component of the field, which in most models can account for up to 10\% of the magnetic energy \cite{LaskyReview}.

If the toroidal field is strong enough the overall shape of the star is prolate. In a young neutron star this leads to an instability which tends to `flip' the star, rapidly leading to inclination angles of $\alpha=\pi/2$, given an initial angle $\alpha\neq 0$, which is the optimal configuration for GW emission in the $l=m=2$ harmonic \cite{Cutler02}. In fact, if the outcome of a core-collapse or a merger is a meta-stable magnetar, rotating with millisecond periods, such an instability will lead to it being a detectable source of GWs for next generation detectors \cite{zhong19}.
The effect of viscosity on secular timescales counteracts the GW torques, and pushes the star back to alignment on longer timescales of hundreds of years.

As the star ages the crust also plays a role, as the field evolves due to the Hall effect, leading to locally enhanced magnetic patches which may lead to detectable deformations in young pulsars \cite{Suvorov16}.

Additionally, as the temperature drops, the interior protons are also expected to become superconducting, leading to an ellipticity of \cite{Cutler02}:
\begin{equation}
\epsilon\approx 10^{-8} \left(\frac{\bar{B}}{10^{12}\mbox{G}}\right)  \left(\frac{\bar{H_c}}{10^{15}\mbox{G}}\right)\, ,
\end{equation}
where $H_c$ is the lower critical field for superconductivity.

\begin{figure}[t]
\begin{center}
\includegraphics[scale=0.7]{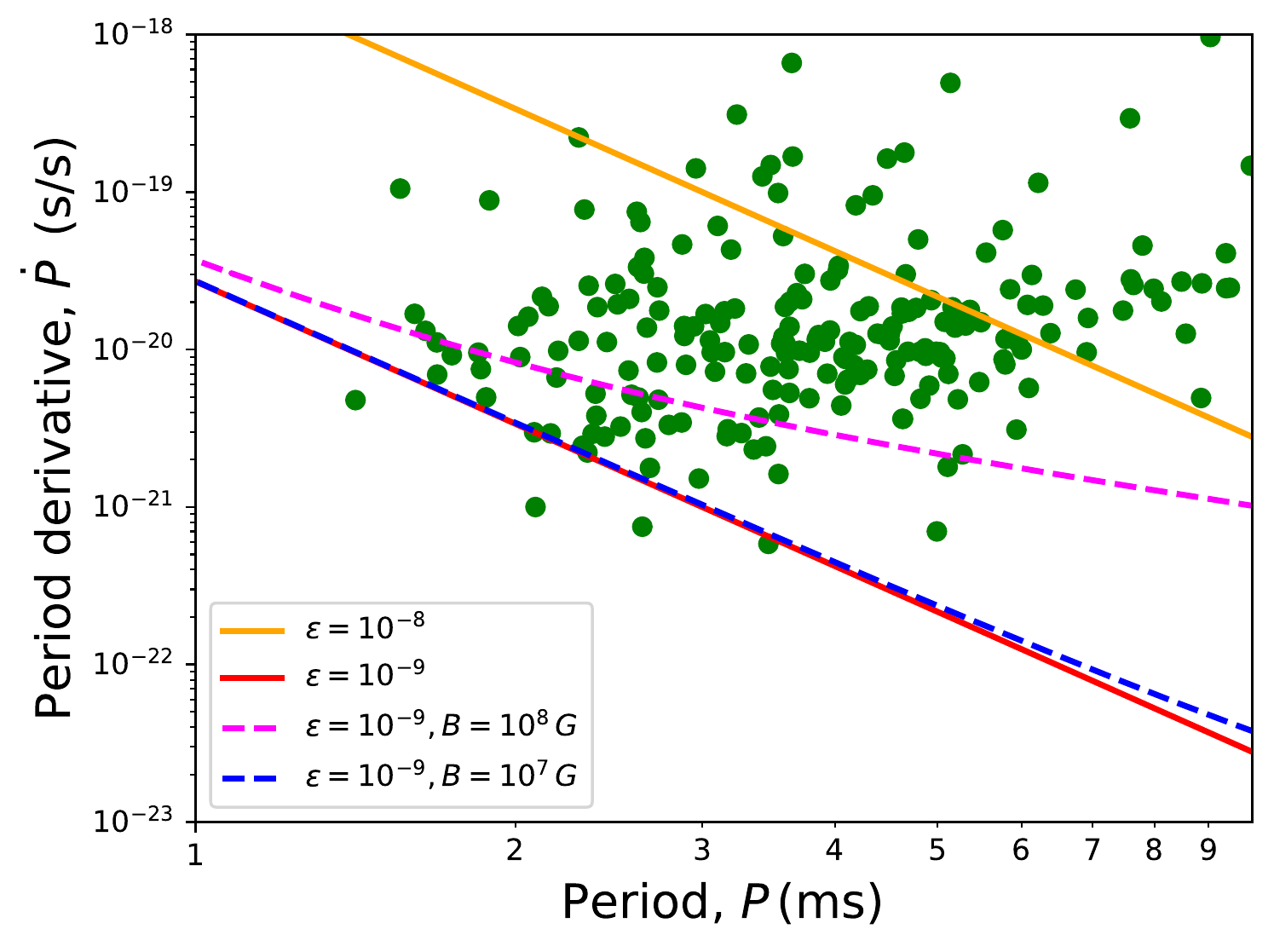}
\end{center}
\caption{Observed periods and period derivatives of millisecond pulsars (the bottom left corner of the standard $P-\dot{P}$ diagram. The solid lines represent gravitational wave spindown models with a given ellipticity, and the dashed lines also include magnetic dipole spindown. As shown in \cite{Woan18} the observed cutoff is consistent with sources being spun down by GW due to a mountain with an ellipticity of $\epsilon\approx 10^{-9}$ and a weak magnetic field of $B\approx 10^{7}$ G. This could signal the presence of a residual ellipticity due to a stronger buried magnetic field in the superconducting core of the stars.}
\label{cutoffppdot}
\end{figure}

Furthermore if the star has undergone episodes of accretion the field may be buried, leading to a lower external dipolar field (inferred from measurements of spin-down rate, if the NS is visible as a pulsar), but allowing for a substantial buried mountain \cite{LaskyReview}, which may be detectable by next generation GW detectors. In fact, the presence of a `minimal' ellipticity due to a buried magnetic field may explain the observed clustering and cutoff, shown in figure (\ref{cutoffppdot}), in the $P-\dot{P}$ diagram of millisecond radio pulsars, that are generally thought to be old NS spun up to millisecond periods by accretion in a binary \cite{Woan18}.

\subsection{Exotic matter and core deformations}

The crust of the NS may not be the only region that can sustain shear deformations and a `mountain'. As we have seen in section \ref{dense}, at high densities, matter may undergo a number of phase transitions in which standard hadronic matter gives way to phases in which quarks are deconfined and possibly paired in the color and flavor degrees of freedom \cite{Alford:2007xm}.
The exact nature of the pairing at NS densities is unknown, as discussed in the previous sections, but in several cases crystalline phases may exist which allow for a sizable mountain. Early studies showed that such a `solid' core could harbour mountains of up to $\epsilon\approx 10^{-4}$, which are within reach of current detectors \cite{LaskyReview}.

A more recent analysis \cite{Glampedakis12} has shown that the presence of strange quark matter and type II superconductivity in the core, leads to color-magnetic mountains, which for a purely poloidal magnetic field have ellipticities of
\begin{equation}
\epsilon^{2SC}\approx 4\times 10^{-6} \bar{B}_{14}\, ,   
\end{equation}
if only the $u$ and $d$ quarks are paired in a two flavor phase, and $\bar{B}_{14}$ is the average field strength in units of $10^{14}$ G, and
\begin{equation}
  \epsilon^{CFL}\approx 1 \times 10^{-5} \bar{B}_{14}\, ,  
\end{equation}
if all three species of quarks are paired in a color-flavor-locked (CFL) phase.
A detection of such a large ellipticity by the next generation of gravitational wave detectors may thus give us a direct indication of the state of matter at high densities in the interior of the NS.

\section{Gravitational wave seismology}
\subsection{Orthogonal oscillation modes and instabilities}

Out of the various oscillation modes of a compact source \cite{Kokkotas:1999bd}, f-modes and r-modes can according to relativistic hydrodynamics become unstable to gravitational wave emission due to the Friedmann-Schutz mechanism \cite{Friedman:1978hf}. This instability increases strongly with the oscillation frequency of these modes, which in turn is roughly proportional to the rotation frequency of the source. R-modes are special since they are in the absence of dissipation unstable at any frequency \cite{Andersson:1997xt}, while viscous dissipation can damp them at sufficiently low frequencies. F-modes in contrast generically only become unstable at frequencies close to the Kepler frequency of the corresponding source. Well established dissipation mechanism in standard neutron stars, like e.g. the Ekman rubbing of the fluid along the solid crust \cite{Lindblom:2000gu}, have been shown to be insufficient to damp r-modes in most millisecond sources. Therefore, r-modes could be present in many astrophysical sources, while f-modes could likely only be present in newborn neutron stars. Exotic phases with enhanced forms of dissipation, like the resonant behavior eq.~(\ref{eq:bulk-viscosity}) of bulk viscosity in quark matter \cite{Alford:2013pma}, or structurally more complicated stars, could damp r-modes in observed sources.

In case a mode becomes unstable its amplitude, conveniently described by a dimensionless amplitude parameter $\alpha$, see eq.~(\ref{eq:r-mode}), grows exponentially. Whereas these modes are orthogonal at infinitesimal amplitude, they mix and couple to other ("daughter"-)modes at finite amplitudes \cite{Bondarescu:2013xwa}. This mechanism can dissipate an increasing amount of energy from the mode as its amplitude grows and can therefore in principle stop the growth and saturate the mode at a finite amplitude $\alpha_{\rm sat}$. Other such saturation mechanisms include the nonlinear enhancement of bulk viscosity at large amplitude \cite{Alford:2010gw}, or the resonant coupling to modes in the crust of the star \cite{Gusakov:2013jwa}. Sources with saturated modes could be emitters of continuous gravitational waves over long time intervals--in case of r-modes in millisecond pulsars potentially for billions of years.

\subsection{f- and r-modes in new-born and young sources}

The evolution of the gravitational wave signal due to unstable modes has generally 3 phases \cite{Owen:1998xg}: (a) a steep rise during the exponential mode growth, (b) a plateau phase as long the amplitude is saturated but the source does not appreciably spin down, yet and (c) a (potentially slow) power law decay as the source spins down. The gravitational wave emission ends as soon as the source spins out of the instability region. At high saturation amplitudes the plateau phase might not be realized, resulting in a short but strong gravitational wave burst, whereas at low amplitude the signal is very stable and hardly changes over long time intervals.

F-modes can very likely only be present in newborn sources originating from a core bounce supernova or neutron star merger, in case the product spins initially close to its Kepler frequency, which is given by the empirical formula $f_K \approx 0.11 \sqrt{GM/R^3}$ in terms of the mass $M$ and radius $R$ of the non-rotating configuration \cite{Paschalidis:2016vmz} and is typically around a kHz. While in Newtonian gravity and for moderate mass sources the f-mode instability region typically extends only down to $0.9 f_K$, in general relativistic simulations of heavy sources it might, depending on the composition, extend to frequencies significantly below $0.8 f_K$ \cite{Paschalidis:2016vmz}. The size and length of the gravitational wave signal depends both on the f-mode growth time and saturation amplitude. In supermassive mergers the typically short live time of the merger product might prevent unstable f-modes to reach sizable amplitudes. The same holds for r-modes which typically have growth times of order seconds. Using mode coupling as saturation mechanism allowed to follow the evolution and predict that f-modes might lead to an observable signal for a sufficiently nearby source \cite{Doneva:2015jaa}. 

In case of newborn sources in supernovae the signal can last until the source spins out of the instability region. While this should still be reasonably quick for f-modes, depending on the saturation amplitude r-modes could be present in young sources for long times \cite{Owen:1998xg}. At sufficiently late times (phase (c) above) the gravitational wave strain of a known source has the interesting property that it is independent of the spin frequency and exclusively determined by its age and its distance \cite{Alford14}
\begin{equation}
h_{0} \approx2.3_{-0.8}^{+3.5}\times10^{-27}\sqrt{\frac{1000\,{\rm y}}{t}}\frac{1\,{\rm Mpc}}{D}\:,\label{eq:numeric-intrinsic-strain}
\end{equation}
where the uncertainty stems mainly from the unknown saturation mechanism.

\begin{figure}[t]
\begin{center}
\includegraphics[scale=0.8]{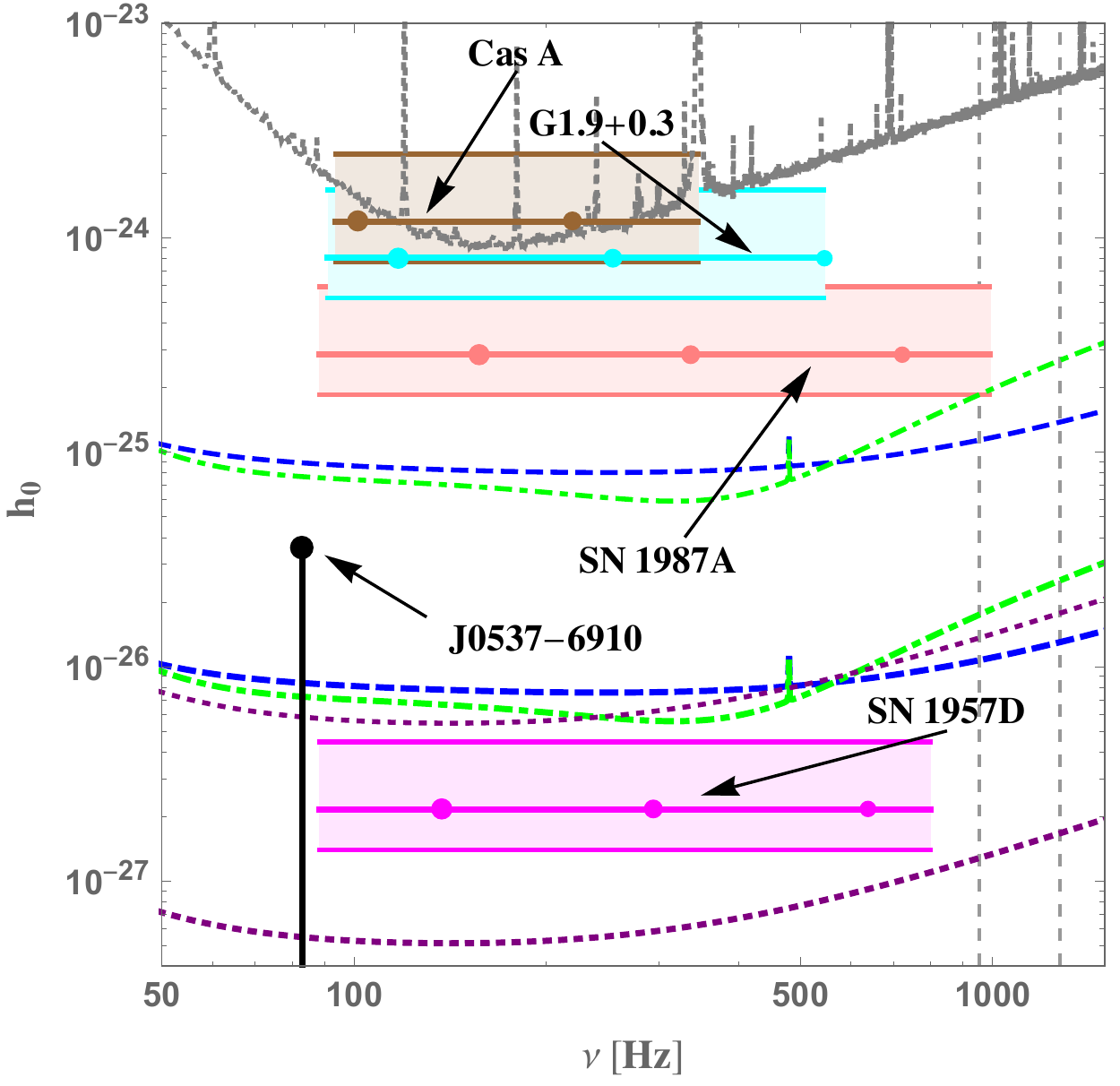}
\end{center}
\caption{Expected intrinsic gravitational wave strain amplitude due to r-modes eq.~(\ref{eq:numeric-intrinsic-strain}) for various young sources \cite{Alford14}, where the error bands stem from the unknown source properties. The frequency range is obtained from an r-mode evolution and  reflects the range of the unknown saturation amplitudes, where the dots denote the corresponding values for different orders of magnitude
from $\alpha_{{\rm sat}}=0.1$ to $10^{-3}$.
These estimates are compared to detector sensitivities of advanced LIGO 
in the standard
mode (dashed) and the neutron star optimized configuration (dot dashed), as well as (dotted) to the Einstein Telescope (ET configuration B from http://www.et-gw.eu/index.php/etsensitivities).
These sensitivities are given both for a known pulsar search with one year of data (thick) and a search for potential sources without timing information using one month of data (thin). the uppermost curve shows the sensitivity of a previous Cas A search.}
\label{fig:young-sources}
\end{figure}

Based on the spindown evolution in the presence of r-modes (discussed below) and based on the observed timing behavior of observed young pulsars, it has been shown that nearly all of them should be stable to r-mode emission. A potential exception is the fastest spinning young pulsar J0537-6910 with a spin frequency of 62 Hz, which might just be in the last stage of its r-mode spindown. Correspondingly, r-mode spindown provides a quantitative explanation for the strikingly low spin frequencies of observed young pulsars \cite{Alford14}. Therefore sources younger than the youngest known pulsars, which have ages around a thousand years, like the neutron star in Cas A or undetected sources associated with recent supernova remnants are very promising for r-mode gravitational wave searches, as shown in figure \ref{fig:young-sources}. In particular the most recent nearby supernova SN1987A is interesting, since there are first indications for thermal emission from the corresponding neutron star \cite{Page:2020gsx}.

\subsection{r-modes in recycled sources}

The progenitors of the millisecond radio pulsars, i.e. the neutron stars in Low Mass X-ray Binaries (LMXBs) that are accreting matter and being spun up, are also an interesting target. 
In these systems the neutron stars all appear to be spinning well below their theoretical breakup frequency, and gravitational wave emission due to unstable r-modes may provide additional spin-down torques that set in once the system enters the unstable region.

\begin{figure}[t]
\begin{center}
\includegraphics[scale=0.7]{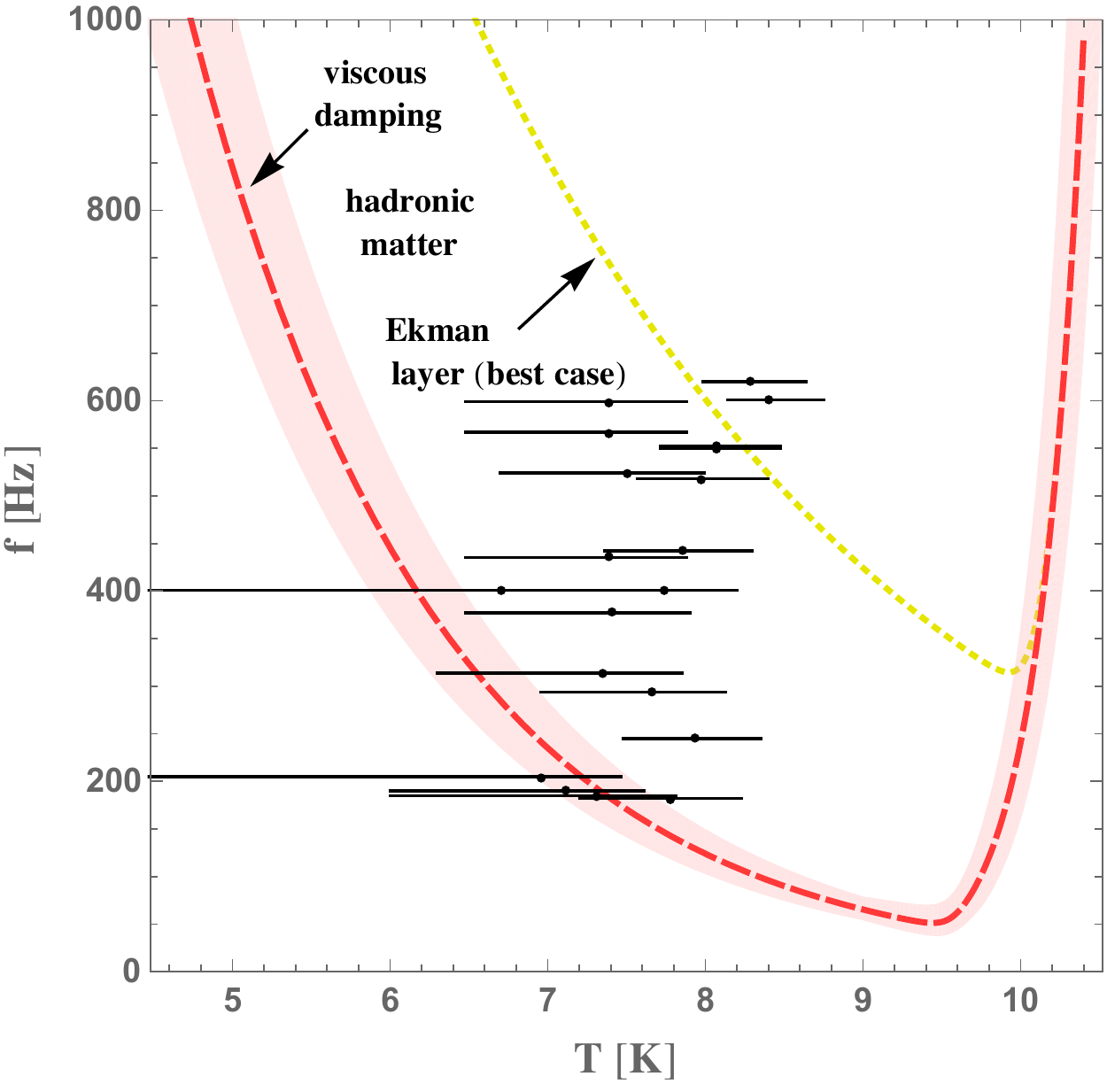}
\end{center}
\caption{Boundary of the r-mode instability
region for a standard hadronic neutron star, taking into account the uncertainties from the micro-physics and unknown source properties, 
compared to thermal data of LMXBs taken from \cite{HaskellDegenaar,Gusakov14}. The main source of damping at high temperatures is bulk viscosity and at low temperatures shear viscosity, but they are insufficient to damp r-modes at high frequencies in the enclosed instability region. Dissipation in an Ekman layer between the crust and the core increases the damping \cite{Lindblom:2000gu}, but, using the appropriate shear viscosity \cite{Shternin:2008es}, it can even in the shown, unrealistic best-case scenario not damp r-modes in the fastest sources \cite{Alford:2013pma}. Despite the large errors in temperature (due to the unknown composition of the outer envelope, which must be modelled to infer the core temperature), it is clear that many systems lie in the unstable region for such a basic NS model.}
\label{fig:instability-regions}
\end{figure}

Fig. \ref{fig:instability-regions} shows the instability region for a `minimal' neutron star composition, while the band estimates the uncertainties. 
However, from the figure it is also clear that such a minimal model is problematic, as several millisecond sources are well inside the instability window. In such a scenario the neutron stars would emit gravitational waves and, unless there is a very strong amplitude-dependent saturation mechanism that can saturate r-modes at low amplitudes, would be rapidly spun down out of the region, making it statistically very unlikely to observe so many \cite{HaskellDegenaar}. The most likely option is thus that additional physics must be at work to damp the r-modes and modify the instability window. The presence of recycled radio pulsars spinning at around 700 Hz also confirms this conclusion, as these systems could not be spun up to such high frequencies if the r-mode would be unstable and reach a sizable amplitude, since the gravitational wave emission would be spinning them down. 

Several additional effects, including superfluid mutual friction, hyperon bulk viscosity and coupling to crustal modes may be at play, and lead to a significantly different instability window at low temperatures, which could be probed by future gravitational wave observations. An intriguing possibility are exotic forms of matter, like quark matter \cite{Alford:2013pma}, where the resonant enhancement of the bulk viscosity at neutron star temperatures could provide the dissipation to completely damp these modes in LMXBs. 

Taking into account that exotic forms of matter or structural complexity increase the damping compared to a minimal neutron star composition and considering the various uncertainties, a general condition for the presence of r-modes has been derived. For a source with an X-ray and/or UV observation or upper bound for the (un-redshifted) surface temperature $T_{s}$, r-modes are absent if it spins with a frequency \cite{Boztepe:2019qmm}
\begin{equation}
f \leq \left(389^{+165}_{-116} \right) {\rm Hz} \; \sqrt{\frac{T_{s}}{10^{5}\,{\rm K}}} \label{eq:frequency-bound}
\end{equation}
While for most millisecond pulsars r-mode emission cannot be excluded based on present X-ray data, the only two millisecond pulsars PSR J0437$-$4715 and PSR J2124$-$3358 with actual temperature measurements are very likely r-mode stable.

Another possibility is that the observed sources are indeed within the instability region, and emit continuous gravitational waves over very long time intervals, but the saturation amplitude for the r-mode is so small that gravitational wave emission does not impact on either the thermal or spin evolution of the stars. However, both the observed spin-down behavior as well as bounds on the thermal X-ray emission, discussed above, set stringent bounds on the r-mode amplitude. These bounds \cite{Boztepe:2019qmm} are by now several orders of magnitude below theoretical predictions for well constrained saturation mechanisms, like mode coupling \cite{Bondarescu:2013xwa}. As discussed, this requires a very strong but yet unidentified source of dissipation within these stars to saturate or even completely damp r-modes in these sources. 

\section{Multi-messenger observations}
\subsection{Gravitational wave driven spin evolution}

As discussed in more detail, below, the detectability of gravitational waves from compact sources is strongly increased for known pulsars, where timing data is available. Since gravitational waves from compact sources generally carry angular momentum, the emission spins down the source. In case sizable deformations or unstable modes are present in observed sources they could easily dominate the spindown of these sources. Therefore, it is important to understand the spindown due to gravitational wave emission. The spindown of a compact source with angular velocity $\Omega$ has the canonical power law form
\begin{equation}
\dot\Omega\sim \Omega^n
\end{equation}
In the case where stellar deformations (`mountains') provide the main contribution to the spindown the power law exponent is $n_{\rm d}=5$, as seen in section 1.4. This is sufficiently different from magnetic dipole emission which typically dominates in the absence of gravitational wave emission and has a canonical exponent of $n_{\rm em}=3$.

In case of r-modes the situation is more complicated since due to the instability the r-mode amplitude has to be saturated by an amplitude-dependent dissipation mechanism, which in addition generally also depends on oscillation frequency and the temperature of the source. This dissipation heats the star so that the evolution of the amplitude, the spin frequency and the thermal evolution, depending also on the cooling mechanism in the source, are coupled. However, except for newborn sources the spindown evolution is much slower than the amplitude and thermal evolution. Parametrizing the cooling power by $P = {\hat P} T^\theta$ and  saturation amplitude $\alpha_{{\rm sat}}=\hat{\alpha}_{{\rm sat}}T^{\beta}\Omega^{\gamma}$ by general power laws (as realized for proposed mechanisms) this results in an effective spindown law of the above form with an effective power law exponent \cite{Alford14}
\begin{equation}
n_{\rm r}=\frac{\left(7\!+\!2\gamma\right)\theta\!+\!2\beta}{\theta\!-\!2\beta} \label{eq:effective-exponent}
\end{equation}
In the prototypical but unrealistic case of a constant saturation amplitude this leads to an exponent of 7, but depending on the saturation mechanism, the spindown law can be rather different. For instance mode coupling leads to an exponent of 4 and for different mechanism it could in principle lie anywhere between 1 and 10 making r-mode searches rather involved.

The detailed r-mode spindown evolution allows us in addition to the above r-mode analysis for LMXBs also to compare timing data of millisecond pulsars, for which there is no thermal data, to theoretical predictions for the boundary of the instability region in timing parameter space \cite{Alford:2013pma}. The corresponding boundary curves have a very similar form in $\dot \Omega$-$\Omega$-space as in $T$-$\Omega$-space, showing that sources with a sufficiently low spindown rate are r-mode stable. However, the observed fast-spinning sources have spindown rates that are high enough that they could all be inside the r-mode instability region, just as in the case of LMXBs. A definite conclusion is in this case so far not possible since other mechanisms could be responsible or dominate the observed spindown.

\subsection{Impact of oscillation modes on the thermal evolution}

As discussed the r-mode evolution strongly affects the thermal evolution as well as electromagnetic observables and therefore indirectly reveals the gravitational emission, even if it is not directly observed. The analysis of r-mode instability regions, discussed before is an example for this. In turn observed temperatures or bounds on temperature impose bounds on the r-mode amplitude in observed sources, since they would otherwise be hotter than what is observed. The bound on the r-mode amplitude for a source with an observed surface temperature $T_s$ takes the form \cite{Boztepe:2019qmm}
\begin{equation}
\alpha_{{\rm sat}} < 1.0\times10^{-9} \left(\frac{T_s}{10\,{\rm eV}}\right)^{2} \left(\frac{500\,{\rm Hz}}{f}\right)^{4} \left(\frac{20 \pi \tilde J M}{1.3\,M_{\odot}}\right)^{-1}  \left(\frac{R}{10\,{\rm km}}\right)^{-2}
\label{eq:numeric-saturation-amplitude-bound}
\end{equation}
In LMXBs where direct temperature measurements are available, this leads to bounds as low as $10^{-8}$ \cite{Mahmoodifar13}. In millisecond pulsars where at present only temperature bounds can be obtained, these nevertheless impose rigorous bounds as low as $\alpha < 3\times 10^{-9}$ \cite{Boztepe:2019qmm} which impose significantly tighter constraints than the spindown of these sources and shows that a detection of a potential r-mode gravitational wave emission from these sources will require third generation detectors.

\section{Continuous GW searches and current bounds}

To date the only detected GW signals involving neutron stars have been inspirals, however the sensitivity of the detectors has been steadily improving and a number of astrophysically significant constraints has been set in continuous wave searches (see \cite{SieniawskaREV} for a recent review), and in the next observational runs of the LIGO/Virgo/KAGRA network there is a very real possibility of detecting these subtle signals. 

\subsection{Present searches}

As we have seen, continuous GWs are expected to be weak, but as they are long lasting signals, longer integration times $T$ can allow to build up signal to noise ratio (SNR) as:
\begin{equation}
SNR\propto h_0 \sqrt{\frac{T}{S}}\, ,    \label{snrligo}
\end{equation}
with $h_0$ the characteristic strain amplitude of the signal, and $S$ the spectral density of the noise in the detectors and the signal frequency.  Consider one of the sources we have discussed in the previous sections, e.g. a galactic NS with a `mountain', rotating with a 1 ms period at 10 kpcs, and with $\epsilon=10^{-7}$. The GW amplitude, from equation (\ref{strainc}) is $h_0\approx 10^{-26}$, and from equation (\ref{snrligo}) we see that to claim a detection with $SNR=5$, given the typical noise spectral density in LIGO for O3, $\sqrt{S}\approx 10^{-23}$ Hz$^{-1/2}$, one needs $T\approx 1$ year. This is a very promising limit, as the observation time needed is comparable to the duration of the O3 run, and as sensitivity increases during the next runs, observations will begin to dig into astrophysically significant parameter space and confirm, or falsify, GW emission models.

In practice, the sensitivity of real world GW searches depends on a trade off between computational costs and accuracy. 
For isolated NSs the parameters describing the signal are the two angles which describe the sky position of the source, right ascension $\alpha$ and declination $\delta$, the rotation frequency of the star $\nu$ (which we have seen, determines, depending on the emission model, the frequency $f_{GW}$ of the GW signal), and the frequency derivative $\dot{\nu}$ which describes its evolution over the observational window (in some cases, for large variations of $\dot{\nu}$ and long observational windows, it may be necessary to consider also $\ddot{\nu}$). 
If the source's position is known, and $\nu$ and $\dot{\nu}$ are accurately determined from EM observations, a {\it targeted} search can be performed. These searches are computationally light, and allow for a larger accuracy and to probe for weaker signals. 
In the case where the parameters are not accurately known, a narrow band search is possible, around the expected values, and while similar in principle is slightly more computationally expensive.

The situation changes if $\nu$ and $\dot{\nu}$ are not known, but the position in the sky of the source is, e.g. if one is searching for CW signals from young neutron stars in supernova remnants, which are not observed as EM pulsars. In this case on can perform a so-called {\it directed} search, scanning over a range of frequencies and frequency derivatives, thus increasing the computational load and decreasing the sensitivity.

Finally, the most computationally expensive (and least sensitive) searches are all-sky {\it blind} searches, where one has to scan the sky for completely unknown signals. Several methods have been developed to deal with this problem and search the data for CW signals from isolated neutron stars, including the $\mathcal{F}$-statistic method, the Hough transform, the 5-vector method, the Band Sampled Data method and the time domain heterodyne method. 
Methods have also been implemented to search for signals from neutron stars in binaries, where the orbital modulation must be accounted for, e.g. the TwoSpect methods, CrossCorr method , Viterbi/$\mathcal{J}$-statistic method  and the Rome narrow-band method, and methods have been devised to detect shorter duration `continuous' signals (see \cite{SieniawskaREV} and references therein for a detailed review of the methods).

\subsection{Electromagnetic constraints}

As mentioned in the previous section, electromagnetic observations can be used to aid GW detections by providing constraints on parameters such as sky position and frequency evolution. However, EM astronomy can reveal valuable information on the emission itself, by constraining the amplitude of the signal and providing indirect evidence that some systems are emitting GW.

\begin{figure}[t]
\includegraphics[scale=0.7]{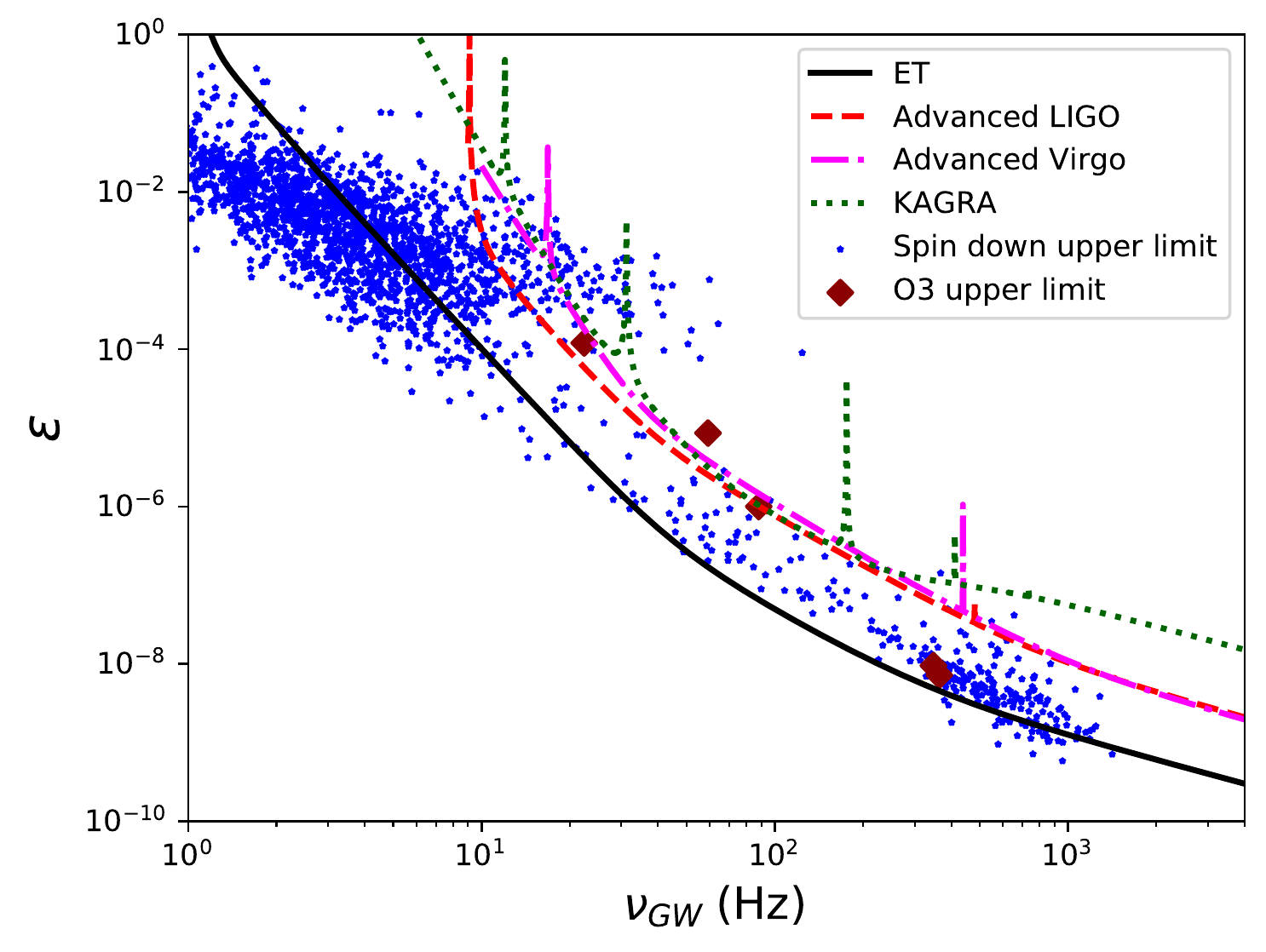}

\caption{Spin-down upper limits on the ellipticity for known pulsars, from the ATNF pulsar catalogue \cite{ATNF}. For reference we include also the sensitivity curves for Advanced LIGO, Advanced Virgo and KAGRA at design sensitivity, assuming a distance of 5 kpc for the source, and an integration time of 1 year. We also show the upper limits obtained by LIGO and Virgo on selected pulsars using O3 data, from \cite{O3a}. Note that some of the pulsars are at a distance of approximately 0.1 kpc, which allows for more stringent upper limits than the sensitivity curves plotted for 5kpc.}
\label{figUL}       
\end{figure}

In the case of CW emission from mountains, measurements of the spin-down rate $\dot{\nu}$ of a pulsar, from timing in the radio, X-ray, or other EM bands, can readily provide us with a useful upper limit on the ellipticity. If, in fact, we assume as an upper limit the case in which the spindown is entirely due to GW emission (which is clearly an upper limit, as we observe the EM emission from the pulsar, thus know it must contribute to the energy budget), i.e. that the star is a {\it gravitar}, from equation (\ref{spindown}) we see that the ellipticity is constrained to be
\begin{equation}
    \epsilon\leq 1.9\times 10^{-5} \left(\frac{\nu}{100\, \mbox{Hz}}\right)^{-5/2}\left(\frac{|\dot{\nu}|}{10^{-10}\,  \mbox{Hz/s}}\right)^{1/2} \left(\frac{I}{10^{45}\, \mbox{g cm$^2$}}\right)^{-1/2}\, .
\end{equation}
For example, for the Crab pulsar, the spin-down upper limit above gives $\epsilon\leq 7.5\times 10^{-4}$, corresponding to a strain of $h_0=1.4\times 10^{-24}$ (see e.g. \cite{NilsBook} for a more detailed derivation). In figure (\ref{figUL}) one can see a plot of the ellipticity spin-down limits for the known pulsars, using data from the ATNF pulsar catalogue\footnote{https://www.atnf.csiro.au/research/pulsar/psrcat/} \cite{ATNF}. As we see from the figure, and will discuss in the next section, the spindown limit is the benchmark against which to test the sensitivity of detectors.

Analogous spindown limits are available for r-modes \cite{Owen:2010ng}. However, taking into account that the same r-mode saturation mechanism should be present in different sources provides in this case even stronger limits
. As discussed before, for sources with X-ray bounds these can impose even much tighter limits showing that r-mode emission from millisecond pulsars is not in reach of present detectors.

EM observations, can, however, provide more than upper limits. We have already seen that in the case of accreting systems, observations of the spin-distribution of the NSs suggest that additional GW torques may be active, as all systems are spinning well below their theoretical maximum frequency, as set by the Keplerian breakup frequency of the star. 

If we assume that the observed spin-period of these systems is set by the equilibrium between GW torques (see eq. \ref{spindown}) and an accretion torque $N_a$ of the form:
\begin{equation}
N_a=\dot{M}\sqrt{(GMR)}\, ,
\end{equation}
where $G$ is the gravitational constant, $M$ and $R$ are the mass and radius of the star and $\dot{M}$ is the accretion rate, which we can infer from the bolometric X-ray flux $L$ of a source at a distance $d$, according to
\begin{equation}
    \dot{M}\approx \frac{4\pi R d^2 F}{GM}\, .
\end{equation}
We can thus obtain a torque balance upper limit:

\begin{eqnarray}
 \epsilon_{tb}&=&2.4\times 10^{-7} \left(\frac{F}{10^{-8} \mbox{ erg cm$^2$ s$^{-1}$}}\right)^{1/2}  \left(\frac{R}{10 \mbox{ km}}\right)^{3/4}  \left(\frac{M}{1.4 M_\odot}\right)^{-1/4}\nonumber\\
 &&\left(\frac{\nu}{100 \mbox{ Hz}}\right)^{-5/2} \left(\frac{d}{1 \mbox{ kpc}}\right)  \left(\frac{I}{10^{45} \mbox{ g cm$^2$}}\right)^{-1}\, .  
\label{torquelimit} \end{eqnarray}
which is the benchmark for the sensitivity that detector must reach to be able to investigate these sources. For transitional pulsars that are not in torque balance the limit could be weaker, though. 

Furthermore, observations of spins and temperatures in LMXBs allow to reconstruct observationally the r-mode instability window, and provide strong evidence that exotic phases, such as hyperons, or dynamcially significant magnetic fields in superconductors, must be present in the interior of the star, given the lack of detections \cite{HaskellDegenaar}.

Individual systems also show evidence for gravitational wave emission, for example PSR J1023+0038, which transitions from accretion phases where it emits X-rays to quiescent phases where radio emission is present. This system has an enhanced spin-down rate during the accretion phase, which suggests that a gravitational mountain is being built \cite{HaskellPRL17} and may be present also in the quiescent phase \cite{BH20}. 

Another interesting system is the young X-ray pulsar PSR J0537-6910. This is the most prolific glitcher currently known, and measurements of the breaking index between glitches suggest an underlying value of $n=7$, consistent with gravitational wave emission from an unstable r-mode driving the spin-down \cite{Andersson0537}. This scenario is theoretically plausible, given our knowledge of the equation of state of dense matter \cite{Andersson0537}, making this system also a target for ongoing GW searches \cite{Fesik:2020tvn}.

Millisecond radio pulsars, of which the accreting NSs in LMXBs are thought to be the progenitors, also reveal valuable information, and an observed cutoff in the P-Pdot diagram (period vs period derivative) can be well modelled in terms of a residual deformation of the stars, due to a buried magnetic field in the stellar interior, which would provide a {\it lower} limit for the ellipticity \cite{Woan18}.

\begin{figure}[t]
\includegraphics[scale=0.7]{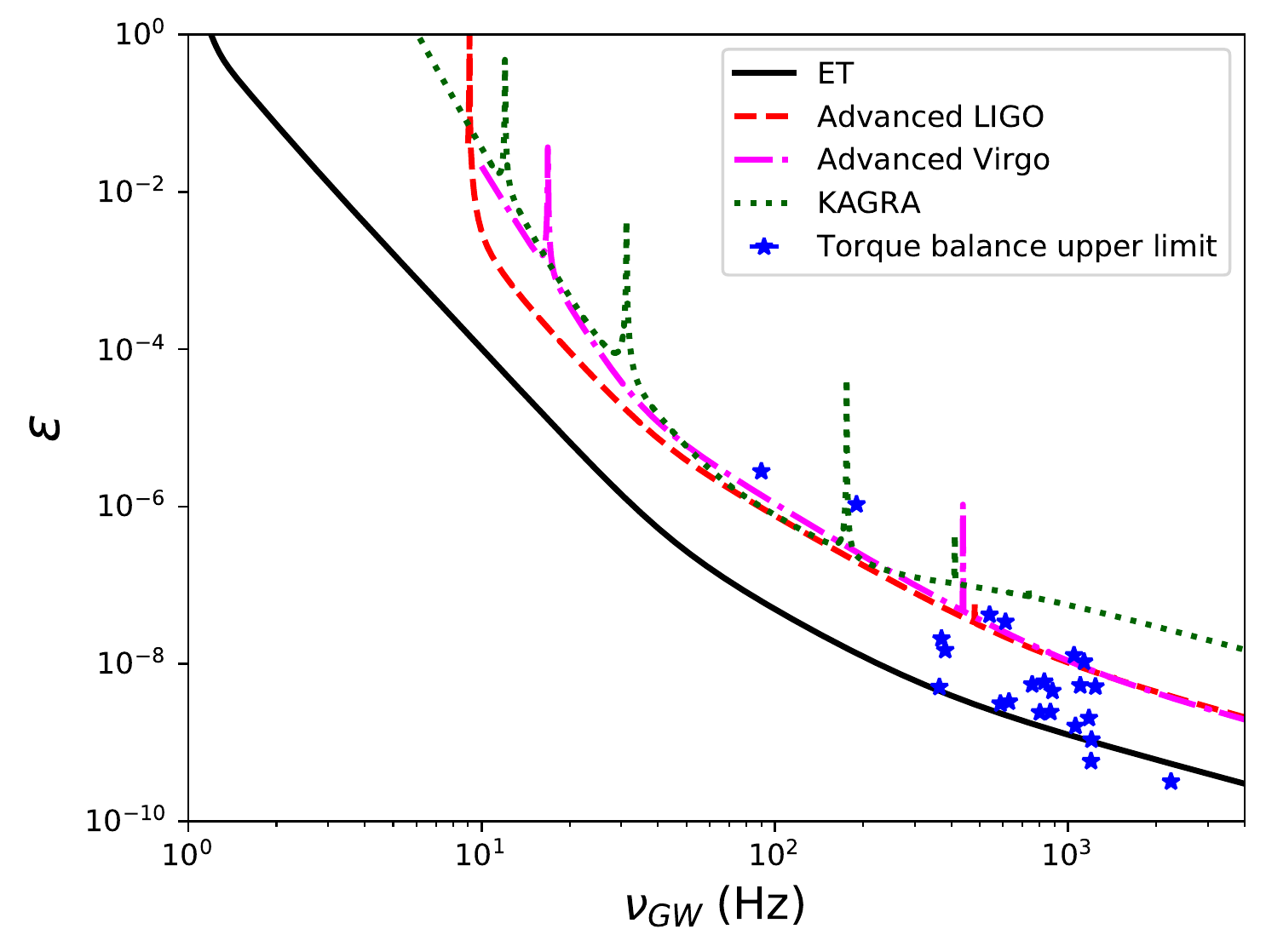}

\caption{Torque balance upper limits on the ellipticity for known accreting sources in LMXBs from \cite{WattsGW}. For reference we include also the sensitivity curves for Advanced LIGO, Advanced Virgo and KAGRA at design sensitivity, assuming a distance of 1 kpc for the source, and an integration time of 1 year.}
\label{figTB}       
\end{figure}

\subsection{Gravitational wave constraints}

At the time of writing no continuous gravitational wave signal has been detected, but, as already mentioned, several pipelines have been developed to search for these signals, and now run on data from all the second observational run of LIGO and Virgo (O2) \cite{SieniawskaREV}, and for some pipelines on parts of data from O3 \cite{O3a}.

The sensitivity of these searches has now reached and beaten the spindown limit in targeted and narrow band searches for a number of pulsars \cite{O3a}, as can be seen in figure (\ref{figUL}), one of which is the Crab pulsar and others, quite notably, millisecond pulsars. In the next observational runs it will thus be possible to probe some of the emission scenarios for millisecond pulsars presented in the previous sections. 

The same is true for searches for GWs from accreting systems, in particular the most luminous X-ray pulsar SCO X-1 \cite{O2SCOX1}. In this case matters are complicated by the fact that the source is not detected as a pulsar, so the frequency is not known. Nevertheless at low frequencies the upper limits are now better than the torque balance limit in (\ref{torquelimit}), so that again astrophysically significant portions of parameter space are being probed. 

There is thus a very real possibility that during the next observational runs of LIGO/Virgo and KAGRA, a continuous wave signal will be detected, opening a new window on neutron star interiors.


\section{Conclusions}

Even though continuous gravitational wave sources have not been observed so far, the future detection of gravitational radiation from isolated neutron stars could provide a wealth of novel information about these extreme objects \cite{Kokkotas:2015gea}. Being emitted by the bulk of the source, they have the potential to probe their otherwise opaque interior. For instance, in case the gravitational waves are emitted by a deformed, rigid quark matter core the observed radiation would literally allow us to see right through the dense hadronic mantle. Similarly, it is generally the dissipation in the bulk of the star that tames unstable oscillation modes and therefore, shapes the corresponding gravitational wave signal. As discussed gravitational waves could be emitted by various different classes of isolated neutron stars at various stages in their evolution. Taking into account that in addition to the few thousand isolated neutron stars known from electromagnetic observations substantially more are expected from galactic evolutionary models, there could be literally millions of sources once the sensitivity of gravitational wave detectors is sufficient to observe them.

Whereas standard hadronic phases very likely cannot sustain deformations that would result in presently detectable gravitational wave emisssion, exotic crystalline phases should be rigid enough that we should see these sources with next generation detectors if they have been sufficiently deformed during their evolution. And while the electromagnetic bounds on the r-mode emission of many recycled sources are by now so tight that they are out of reach of current gravitational wave detectors, several young sources are very promising. Present gravitational wave searches already set important bounds on the continuous emission and several of them already beat the spindown limit for various sources. Therefore they start to probe the composition of neutron stars and even a non-detection can significantly limit the presence of exotic forms of matter inside. Combined with multi-messenger electromagnetic observations such searches have the potential to markedly improve our understanding of neutron stars in the near future.

\begin{acknowledgement}
K.S. has been supported by the Turkish Research Council (TÜBITAK) via projects 117F312 and 119F073.

B.H. has been supported by the National Science Center, Poland (NCN) via grants OPUS 2019/33/B/ST9/00942, OPUS 2018/29/B/ST9/02013 and SONATA BIS 2015/18/E/ST9/00577.
\end{acknowledgement}
%

%
%


\bibliographystyle{spphys}
\bibliography{CWrev}{}


\end{document}